\pdfoutput=1

\documentclass[%
 reprint,
 amsmath, amssymb, superscriptaddress,
 aps, prx
]{revtex4-2}

\usepackage{graphicx}
\usepackage{dcolumn}
\usepackage{bm}
\usepackage{bbm}
\usepackage{physics}
\usepackage{siunitx}
\usepackage{floatrow}
\usepackage{hyperref}
\usepackage{qcircuit}

\begin{document}

\title{Robust photon-mediated entangling gates between quantum dot spin qubits}

\author{Ada Warren}
\affiliation{Department of Physics, Virginia Tech, Blacksburg, VA 24061, USA}
\author{Utkan G\"ung\"ord\"u}
\altaffiliation{Current address: Laboratory for Physical Sciences, College Park, Maryland 20740, USA}
\affiliation{Department of Physics, University of Maryland Baltimore County, Baltimore, MD 21250, USA}
\author{J. P. Kestner}
\affiliation{Department of Physics, University of Maryland Baltimore County, Baltimore, MD 21250, USA}
\author{Edwin Barnes}
\author{Sophia E. Economou}
\affiliation{Department of Physics, Virginia Tech, Blacksburg, VA 24061, USA}

\begin{abstract}
	Significant experimental advances in single-electron silicon spin qubits have opened the possibility of realizing long-range entangling gates mediated by microwave photons. Recently proposed iSWAP gates, however, require tuning qubit energies into resonance and have limited fidelity due to charge noise. We present a novel photon-mediated cross-resonance gate that is consistent with realistic experimental capabilities and requires no resonant tuning. Furthermore, we propose gate sequences capable of suppressing errors due to quasistatic noise for both the cross-resonance and iSWAP gates.
\end{abstract}
\maketitle

\section{Introduction}\label{sec:intro}

Advanced semiconductor fabrication techniques, long coherence times \cite{Veldhorst2014}, and high-fidelity single \cite{Takeda2016, Yoneda2018, Kawakami2016} and two-qubit \cite{Russ2018, Zajac2018, Sigillito2019} gates have positioned solid-state electronic spin qubits as one of the most favorable candidates for quantum information processing \cite{Kloeffel2013, Hanson2007, Awschalom2013, Srinivasa2016, Volk2019, Mortemousque2020, Mills2019}. Recent breakthrough experimental work has demonstrated a coherent interface between individual electron spins in double quantum dots (DQDs) and photons in superconducting microwave resonators \cite{Viennot2015, Mi2018, Samkharadze2018, Landig2018}. By electrically coupling a plunger gate above one dot to a probe in the resonator, and coupling the spin and position degrees of freedom with a nearby micromagnet, Refs.~\cite{Viennot2015, Mi2018, Samkharadze2018, Landig2018} were able to realize a spin-charge hybridized qubit which inherits a long coherence time from its spin-like character and strong coupling to the resonator mediated by the position degree of freedom \cite{Beaudoin2016, Benito2017, Mi2018}. Building on this, recent theoretical work has investigated the use of this spin-photon interface to realize long-range spin-spin entangling gates mediated by resonator photons \cite{Warren2019, Benito2019}. This opens the possibility for large, scalable quantum information processors based on DQD electronic spins. The iSWAP proposed in Refs.~\cite{Warren2019, Benito2019}, however, requires qubits to be tuned into resonance, which can be challenging in some architectures \cite{Borjans2020} and may be impractical for collections of many spins coupled to a common resonator. Additionally, spin-charge hybridization results in susceptibility of the qubit to charge noise, limiting achievable gate fidelities \cite{Mi2018, Warren2019, Benito2019}.

In this paper, we present a protocol for a novel entangling gate in systems of DQDs coupled by microwave resonators: a cross-resonance gate that is locally equivalent to a CNOT and similar to gates used in superconducting transmon qubit systems \cite{Rigetti2010, Paraoanu2006}. We also propose two protocols for suppressing charge noise, including a nested gate sequence based on fast, dynamically corrected single-qubit gates \cite{Zeng2018} which is also able to suppress errors due to quasistatic charge noise for the previously introduced resonant iSWAP. We find that these gate sequences substantially reduce gate infidelity due to quasistatic charge noise.

The paper is organized as follows. In Sec.~\ref{sec:ham}, we introduce the resonator-DQD Hamiltonian and define notation. Our cross-resonance gate protocol is presented in Sec.~\ref{sec:CRgate}. We include quasistatic charge noise in our model and present dynamically corrected iSWAP and cross-resonance gates in Sec.~\ref{sec:DCGs}. We conclude in Sec.~\ref{sec:conclusion}. 

\section{Hamiltonian}\label{sec:ham}

As in Refs. \cite{Warren2019, Benito2019}, we consider a system of several gate-defined DQDs, each tuned to the single-electron regime and capacitively coupled with coupling constant $g_i^{AC}$ to a common microwave resonator mode with frequency $\omega_r$. The inter-dot tunneling constants $t_{c i}$ and detunings $\epsilon_i$ of each DQD are independently electrically tunable, and we explicitly include microwave-frequency electric drive of the detunings with drive frequencies $\omega_i^d$ and envelopes $\tilde{\Omega}_i(t)$. Micromagnets near each DQD, along with an external magnetic field, create an inhomogeneous magnetic field in the vicinity of each DQD. At each DQD, the longitudinal average magnetic field gives rise to a Zeeman splitting $\omega_i^z$ between the DQD electron spin states, while the magnetic field gradients, which for simplicity we take to be transverse, couple the spin and position degrees of freedom of the electrons with coupling strengths $g_i^x$. The system can then be described with the Hamiltonian ($\hbar=1$)
\begin{align}
	\tilde{H}(t) &= \tilde{H}_0 + \tilde{H}_I + \tilde{H}_{dr}(t), \label{eq:FullHam} \\
	\tilde{H}_0 &= \omega_r a^\dag a + \sum_i \qty(\frac{1}{2}\epsilon_i\tilde{\tau}_i^z + t_{c i}\tilde{\tau}_i^x + \frac{1}{2}\omega_i^z \tilde{\sigma}_i^z + g_i^x \tilde{\sigma}_i^x \tilde{\tau}_i^z), \nonumber \\
	\tilde{H}_I &= \sum_i g_i^{AC} \qty(a^\dag + a) \tilde{\tau}_i^z, \nonumber \\
	\tilde{H}_{dr}(t) &= \sum_i \tilde{\Omega}_i(t) \cos(\omega_i^d t) \tilde{\tau}_i^z, \nonumber
\end{align}
where $\tilde{\sigma}_i^k$ and $\tilde{\tau}_i^k$ for $k \in \qty{x, y, z}$ are the spin and position Pauli matrices of the $i$th DQD electron.

To proceed, we transform to the spin-orbit hybridized eigenbasis of $\tilde{H}_0$, in which
\[\tilde{H}_0 = \omega_r a^\dag a + \sum_i \qty(\frac{1}{2}\omega_i^{\tau} \tau_i^z + \frac{1}{2}\omega_i^{\sigma} \sigma_i^z),\]
where $\tau_i^k, \sigma_i^k$ are the transformed Pauli matrices in the new basis and $\omega_i^\tau > \omega_i^\sigma$. We assume $2 t_{c i} > \omega_i^z$ so that the low-energy $\ev{\tau_i^z} = -1$ subspace of each DQD constitutes a qubit which is largely spin-like in character.

In the dispersive regime $1 \gg g_i^{AC} / \abs{\omega_r - \omega_i^\sigma} \gg g_i^{AC} / \abs{\omega_i^\tau - \omega_r}$, we can use a Schrieffer-Wolff transformation to remove the couplings between the DQDs and the resonator to leading order \cite{Schrieffer1966}. Next, we take the empty-cavity limit and assume the microwave drives are weak and well-detuned from both the resonator frequency $\omega_r$ and the larger DQD transition frequencies $\omega_i^\tau$ so that we can neglect transitions out of the $\ev{a^\dag a} = 0, \ev{\tau_i^z} = -1$ subspace. Projecting onto this subspace, we obtain an effective Hamiltonian describing the dynamics of our low-energy qubits \cite{supplement}, in which we see the appearance of long-range qubit-qubit interactions mediated by the resonator mode:
\begin{align*}
	H(t) &= \sum_i \qty(\frac{1}{2} \omega_i \sigma_i^z + \cos(\omega_i^d t)\qty(\Omega_i^z(t) \sigma_i^z + \Omega_i^x(t) \sigma_i^x)) \\
	&\hspace{2.5mm}- \sum_{i<j}J_{ij}\sigma_i^x \sigma_j^x.
\end{align*}

\section{Cross-resonance gate}\label{sec:CRgate}

To arrive at the cross-resonance gate, we focus on a system of two DQD qubits coupled via a resonator. Here, qubit 2 acts as the target and remains undriven, while the control, qubit 1, is driven. For simplicity, we assume a square-envelope microwave pulse, although higher gate performance can be achieved with pulse shaping techniques \cite{Calderon-Vargas2019}. During the pulse, the effective 2-qubit lab-frame Hamiltonian takes the form
\[H(t) = \frac{1}{2}\omega_1 \sigma_1^z + \cos(\omega_1^d t) \qty(\Omega_1^z \sigma_1^z + \Omega_1^x \sigma_1^x) + \frac{1}{2}\omega_2 \sigma_2^z - J \sigma_1^x \sigma_2^x.\]

We proceed following Ref.~\cite{Rigetti2010}, moving into the doubly-rotating frame defined by the transformation $U_a = \exp[-i t \omega_1^d\qty(\sigma_1^z + \sigma_2^z)/2]$. Then, after making the rotating wave approximation (RWA) to eliminate single-qubit terms oscillating at multiples of $\omega_1^d$, we diagonalize the remaining single-qubit terms with the time-independent transformation $U_b = \exp[-i \chi \sigma_1^y / 2]$, where $\chi = \arctan(\Omega_1^x / \delta_1)$ and $\delta_i = \omega_i - \omega_1^d$. In this diagonalized doubly-rotating frame (DDF), the Hamiltonian becomes
\begin{align*}
	H_{DDF} &= \frac{1}{2}\eta \sigma_1^z + \frac{1}{2} \delta_2 \sigma_1^z - J \qty(\cos(\omega_1^d t) \sigma_2^x - \sin(\omega_1^d t) \sigma_2^y) \\
	&\hspace{2.5mm}\times\qty(\cos(\omega_1^d t) (\sin(\chi) \sigma_1^z + \cos(\chi) \sigma_1^x) - \sin(\omega_1^d t) \sigma_1^y),
\end{align*}
where $\eta = \sqrt{\delta_1^2 + \qty(\Omega_1^x)^2}$. Next, we eliminate single-qubit terms with another time-dependent transformation $U_c = \exp[-i t \qty(\eta \sigma_1^z + \delta_2 \sigma_2^z) / 2]$. In this quadruply-rotating frame, we get the Hamiltonian
\begin{align*}
	H_{QF} &= -J [\cos(\omega_2 t) \sigma_2^x - \sin(\omega_2 t) \sigma_2^y]\times[\cos(\omega_1^d t)(\sin(\chi) \sigma_1^z \\
	&\hspace{3mm} + \cos(\chi) \qty(\cos(\eta t)\sigma_1^x - \sin(\eta t)\sigma_1^y)) \\
	&\hspace{3mm} - \sin(\omega_1^d t)\qty(\cos(\eta t)\sigma_1^y + \sin(\eta t)\sigma_1^x)].
\end{align*}

Generically, all terms in this frame oscillate rapidly. However, by choosing a microwave pulse resonant with our target qubit so that $\omega_1^d = \omega_2$, and assuming $\eta \gg J$ so that we can again make the RWA and neglect remaining oscillating terms, we arrive at the time-independent Hamiltonian
\[H_{QF} \approx -\frac{1}{2}\tilde{J}\sigma_1^z\sigma_2^x,\]
where we have defined $\tilde{J} = J\sin(\chi) = J \Omega_1^x/\eta$. Up to local operations, then, this microwave pulse produces a controlled $x$-rotation of the target qubit. In particular, when $\tilde{J} t = \pi/2$, we get a local CNOT equivalent \cite{Rigetti2010}.

Notably, for a given effective coupling $J$, this cross-resonance CNOT is always slower than the previously-introduced resonant iSWAP by a factor of $\Omega_1^x/\eta$. This factor is small when the qubit-qubit detuning $\Delta = \omega_1 - \omega_2$ is large compared to accessible drive strengths. Unlike the iSWAP, however, there is no need for $\Delta$ to be made small relative to $J$. Additionally, since qubits never need to be tuned into or out of resonance, the DQDs can remain at the $\epsilon_i = 0$ sweet spot, allowing for decreased sensitivity to electrical fluctuations \cite{Benito2017}.

To verify our effective model, we simulate the unitary time evolution of the full 2-DQD system, including orbital degrees of freedom and a single resonator mode truncated to 10 photonic states. We focus on systems with realistic static parameters taken from Ref.~\cite{Mi2018}. To ensure suppression of entangling interactions in the absence of microwave driving, we choose Zeeman splittings $\omega_i^z$ such that the qubits are well-detuned from one another. We choose drive amplitudes consistent with reported EDSR Rabi frequencies in single-electron silicon quantum dots \cite{Yoneda2018, Neumann2015}, and drive frequencies such that $\omega_1^d = \omega_2$, following our analytical expressions.

\begin{figure}
	\input{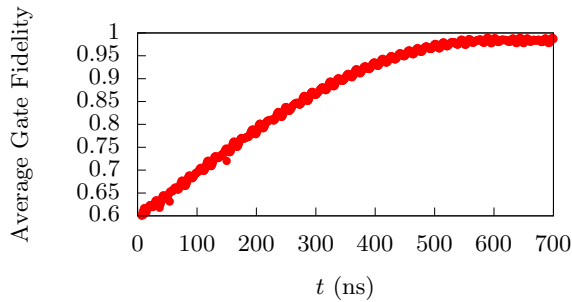}
	\caption{Average fidelity of the cross-resonance CNOT gate, maximized at each time step over local rotations. Here, we have chosen $\omega_r = \SI{6}{\giga\hertz}$, $\omega_1^z = \SI{5.96}{\giga\hertz}$, $\omega_2^z = \SI{5.94}{\giga\hertz}$, $\epsilon_1 = \epsilon_2 = 0$, $2 t_{c 1} = 2 t_{c 2} = \SI{7}{\giga\hertz}$, $g_1^{AC} = g_2^{AC} = \SI{40}{\mega\hertz}$, and $g_1^x = g_2^x = \SI{200}{\mega\hertz}$. Starting at time $t = 0$, qubit 1 is subject to a microwave drive with $\tilde{\Omega}_1$ chosen so that $\Omega_1^x = \SI{15}{\mega\hertz}$ and with drive frequency $\omega_1^d = \SI{5.9059}{\giga\hertz}$. At time $t = \SI{590}{\nano\second}$, the microwave drive is stopped. The final gate fidelity is between $98\%$ and $99\%$, with remaining infidelity primarily due to leakage to excited resonator states.}
	\label{fig:CNOT_fid}
\end{figure}

We numerically solve the Schr\"odinger equation with the Hamiltonian in Eq. (\ref{eq:FullHam}) in the eigenbasis of $\tilde{H}_0$. At each time step, we use the evolved states to compute density operators of the full system, and then trace out the higher-energy DQD and resonator degrees of freedom to obtain reduced density operators describing the qubit evolution. We then compute state-averaged gate fidelities according to the formula in Ref.~\cite{Nielsen2002}: $\bar{F}(\mathcal{E}, U) = \frac{1}{5} + \frac{1}{80}\sum_{j,k=\mathbbm{1},x,y,z} \tr(U \sigma_{1}^j\sigma_{2}^k U^\dag \mathcal{E}(\sigma_{1}^j\sigma_{2}^k))$ where $U$ is our target gate and $\mathcal{E}(\rho)$ is the quantum process describing the noisy time evolution of the qubits. We plot average gate fidelities relative to a perfect CNOT in Fig.~\ref{fig:CNOT_fid}. We find that with realistic parameters, we are able to realize a local CNOT equivalent with $99\%$ fidelity in $\SI{590}{\nano\second}$, with remaining infidelity largely due to leakage to excited resonator states.

\section{Dynamically corrected gates}\label{sec:DCGs}

\subsection{Corrected cross-resonance gate}

We can model the effects of quasistatic charge noise on our cross-resonance gate by substituting $\epsilon_i \to \epsilon_i + \delta\epsilon_i$ and $t_{c i} \to t_{c i} + \delta t_{c i}$ for the detunings and tunnel couplings respectively, where $\delta\epsilon_i$ and $\delta t_{c i}$ are Gaussian-distributed random variables with standard deviations $\sigma_\epsilon$ and $\sigma_t$, respectively. In terms of the low-energy dynamics, the effect of these substitutions is random shifts in the qubit splittings $\omega_i$, drive strengths $\Omega_i^x$, and effective qubit-qubit interaction $J$. The noisy lab-frame Hamiltonian is
\begin{align*}
	H(t) &= \frac{1}{2}\qty(\omega_1 + \delta\omega_1)\sigma_1^z + \frac{1}{2}\qty(\omega_2 + \delta\omega_2)\sigma_2^z - \qty(J + \delta J)\sigma_1^x\sigma_2^x \\
	&\hspace{3mm} + \cos(\omega_2 t)\qty(\qty(\Omega_1^z + \delta\Omega_1^z)\sigma_1^z + \qty(\Omega_1^x + \delta\Omega_1^x)\sigma_1^x).
\end{align*}

As in the noiseless case, we move into the frame rotating with the drive and diagonalize single-qubit terms. Then, discarding all of the same rapidly-oscillating terms as before, we arrive at the noisy DDF Hamiltonian
\begin{equation}
	\label{eq:noisy_DDF}
	H_{DDF} = \frac{1}{2}\qty(\eta + \delta\eta) \sigma_1^z + \frac{1}{2}\delta\omega_2 \sigma_2^z - \frac{1}{2}\qty(\tilde{J} - \delta\tilde{J}) \sigma_1^z \sigma_2^x.
\end{equation}
Note that, because of variations $\delta\omega_1$ and $\delta\Omega_x$, this is actually not the same DDF as in the noiseless case, but is related to it by an additional $\sigma_1^y$ rotation of angle $\delta\chi$.

We can compute the average fidelity $\bar{F}$ of the noisy gate $U^{(1)}_\phi = \mathcal{T} \exp(-i\int_0^{\phi/\tilde{J}} H_{DDF}(t')\dd{t'})$ relative to the noiseless gate. Expanding to lowest order in each of the shifted parameters, the average gate fidelity for a cross-resonance CNOT ($\phi = \pi/2$) is
\[\bar{F} \approx 1 - \frac{\pi^2}{20}\qty(\frac{\delta\eta}{\tilde{J}})^2 - \frac{2}{5}\qty(\frac{\delta\omega_2}{\tilde{J}})^2 - \frac{\pi^2}{20}\qty(\frac{\delta\tilde{J}}{\tilde{J}})^2 - \frac{2}{5}\delta\chi^2.\]

\begin{figure}
	\input{sensitivities-tex}
	\caption{The sensitivities $\qty(\partial x = \sum_i\abs{\pdv{x}{t_{c i}}})$ of various system parameters at the charge degeneracy sweet spot $\qty(\epsilon_1 = \epsilon_2 = 0)$ \cite{Benito2017} with $\omega_r = \SI{6}{\giga\hertz}$, $t_{c 1} = t_{c 2} = t_c$, $g_1^{AC} = g_2^{AC} = \SI{40}{\mega\hertz}$, and $g_1^x = g_2^x = \SI{200}{\mega\hertz}$ \textbf{(a)} for a cross-resonance CNOT with $\omega_1 = \SI{5.96}{\giga\hertz}, \omega_2 = \SI{5.94}{\giga\hertz}$, and $\Omega_1^x = \SI{20}{\mega\hertz}$. \textbf{(b)} for a resonant iSWAP with $\omega_1 = \omega_2 = \SI{5.95}{\giga\hertz}$.}
	\label{fig:sensitivities}
\end{figure}

The sensitivities of various cross-resonance gate parameters to charge noise are plotted in Fig.~\ref{fig:sensitivities}a for some realistic system parameters. In this regime, we see that $\eta$ and $\omega_2$ are much more sensitive to electrical fluctuations than $\chi$ or $\tilde{J}$. For this reason, we neglect errors due to $\delta\tilde{J}$ and $\delta\chi$, and focus our efforts instead on correcting the larger errors. Notably, there is a sweet spot at which $\eta$ is first-order insensitive to charge noise fluctuations. This can be understood from competing effects of $\delta\omega_1$ and $\delta\Omega_1^x$. For $2 t_{c 1} > \omega_1^z$, spin-charge hybridization decreases $\omega_1$. Thus, when fluctuations increase spin-charge hybridization, $\delta\omega_1 < 0$. Meanwhile, the drive strength felt by the qubit increases, so $\Omega_1^x \delta\Omega_1^x > 0$. By choosing $\Delta > 0$ and using an appropriate drive amplitude, then, we can engineer a situation in which $\delta\eta \approx \frac{1}{\eta}\qty(\Delta \delta\omega_1 + \Omega_1^x\delta\Omega_1^x) = 0$.

The noisy 2-qubit Hamiltonian in Eq. (\ref{eq:noisy_DDF}) belongs to an $\mathfrak{su}(2)\oplus\mathfrak{u}(1)$ subalgebra of $\mathfrak{su}(4)$, with $\mathfrak{su}(2)$ generators $\{\sigma_1^z\sigma_2^x, \sigma_1^z\sigma_2^y, \sigma_2^z\}$, all of which commute with the $\mathfrak{u}(1)$ generator $\sigma_1^z$. While the $\delta\eta$ error commutes with the $\sigma_1^z\sigma_2^x$ generator and can be eliminated with a simple $\pi$-pulse (as we discuss below), the $\delta\omega_2$ error anticommutes with it, and requires more nontrivial error correction. 

\begin{figure*}
	\begin{minipage}[c]{0.33\textwidth}
	    \textbf{(a)}
	    \raisebox{-0.9\height}{\input{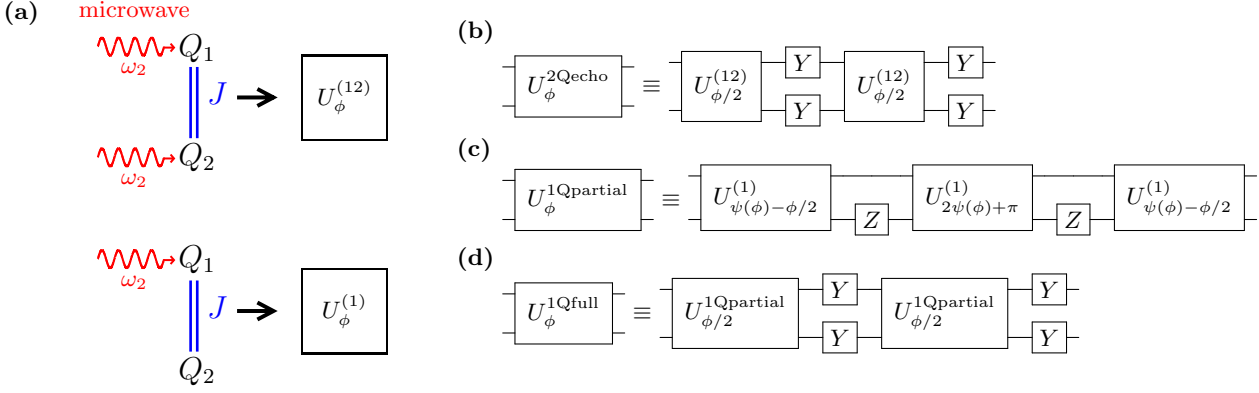}}
	\end{minipage}
	\begin{minipage}[c]{0.66\textwidth}
		\flushleft
		\textbf{(b)}\\
		\hspace{5mm}
		\begin{minipage}[c]{0.1\textwidth}
			\Qcircuit @C=0.5em @R=0.7em {
				& \multigate{1}{U_\phi^{\text{2Qecho}}} & \qw \\
				& \ghost{U_\phi^{\text{2Qecho}}} & \qw
			}
		\end{minipage}
		$ \equiv$
		\begin{minipage}[c]{0.85\textwidth}
			\Qcircuit @C=0.5em @R=0.7em {
				& \multigate{1}{U_{\phi/2}^{(12)}} & \qw & \gate{Y} & \qw & \multigate{1}{U_{\phi/2}^{(12)}} & \qw & \gate{Y} & \qw \\
				& \ghost{U_{\phi/2}^{(12)}} & \qw & \gate{Y} & \qw & \ghost{U_{\phi/2}^{(12)}} & \qw & \gate{Y} & \qw
			}
		\end{minipage}\\~\\
		\textbf{(c)}\\
		\hspace{5mm}
		\begin{minipage}[c]{0.1\textwidth}
			\Qcircuit @C=0.5em @R=0.7em {
				& \multigate{1}{U_\phi^{\text{1Qpartial}}} & \qw \\
				& \ghost{U_\phi^{\text{1Qpartial}}} & \qw
			}
		\end{minipage}
		$ \equiv$
		\begin{minipage}[c]{0.85\textwidth}
			\Qcircuit @C=0.5em @R=0.7em {
				& \multigate{1}{U_{\psi(\phi) - \phi/2}^{(1)}} & \qw & \qw & \qw & \multigate{1}{U_{2\psi(\phi) + \pi}^{(1)}} & \qw & \qw & \qw & \multigate{1}{U_{\psi(\phi) - \phi/2}^{(1)}} & \qw \\
				& \ghost{U_{\psi(\phi) - \phi/2}^{(1)}} & \qw & \gate{Z} & \qw & \ghost{U_{2\psi(\phi) + \pi}^{(1)}} & \qw & \gate{Z} & \qw & \ghost{U_{\psi(\phi) - \phi/2}^{(1)}} & \qw
			}
		\end{minipage} \\~\\
		\textbf{(d)}\\
		\hspace{5mm}
		\begin{minipage}[c]{0.1\textwidth}
			\Qcircuit @C=0.5em @R=0.7em {
				& \multigate{1}{U_\phi^{\text{1Qfull}}} & \qw \\
				& \ghost{U_\phi^{\text{1Qfull}}} & \qw
			}
		\end{minipage}
		$ \equiv$
		\begin{minipage}[c]{0.85\textwidth}
			\Qcircuit @C=0.5em @R=0.7em {
				& \multigate{1}{U_{\phi/2}^{\text{1Qpartial}}} & \qw & \gate{Y} & \qw & \multigate{1}{U_{\phi/2}^{\text{1Qpartial}}} & \qw & \gate{Y} & \qw \\
				& \ghost{U_{\phi/2}^{\text{1Qpartial}}} & \qw & \gate{Y} & \qw & \ghost{U_{\phi/2}^{\text{1Qpartial}}} & \qw & \gate{Y} & \qw
			}
		\end{minipage}
	\end{minipage}
	\caption{\textbf{(a)} Diagram for the different drive schemes used to generate a cross-resonance CNOT. In either case, we have two qubits coupled via a microwave resonator with coupling strength $J$. For the gate $U_\phi^{(12)}$, which is used in 2Qecho, both the control and target qubit are driven at the target qubit transition frequency. For the gate $U_\phi^{(1)}$, which is used in 1Qpartial and 1Qfull, only the control qubit is driven. \textbf{(b)} Circuit diagram for the 2Qecho corrected CNOT. \textbf{(c)} Circuit diagram for the 1Qpartial corrected CNOT or iSWAP, which corrects only the non-commuting errors. \textbf{(d)} Circuit diagram for the 1Qfull corrected CNOT or iSWAP, which corrects both the commuting and non-commuting errors.}
	\label{fig:gatecircuits}
\end{figure*}

One option for suppressing the $\delta\omega_2$ error, inspired by techniques used in superconducting qubits \cite{Sheldon2016, Sundaresan2020}, is the addition of a microwave-frequency drive applied to the target qubit concurrently with the cross-resonance drive applied to the control qubit (Fig.~\ref{fig:gatecircuits}a). By driving the target qubit at its own transition frequency and in-phase with the cross-resonance drive, we introduce a large $(\Omega_2^x + \delta\Omega_2^x)\sigma_2^x$ term to the noisy $H_{DDF}$. As this new term commutes with our desired $\sigma_1^z\sigma_2^x$ generator, but anticommutes with $\sigma_2^z$, this additional driving actively suppresses the $\delta\omega_2$ error without interfering with entanglement generation. This comes at the expense of introducing a new $\delta\Omega_2^x$ error. However, this can be eliminated, along with the $\delta\eta$ error, by a $\pi$ rotation about the $\sigma_i^y$ axis on each qubit. The entire gate sequence, with simultaneous drive on both qubits and single-qubit echo pulses, we refer to as ``2Qecho," and is shown in Fig.~\ref{fig:gatecircuits}b. For $\Omega_2^x \gg \tilde{J}$, the average gate fidelity to lowest order in the presence of this cancellation pulse is
\[\bar{F}_{\text{2Qecho}} \approx 1 - \frac{2}{5}\qty(\frac{\delta\omega_2}{\Omega_2^x})^2 - \frac{\pi^2}{20}\qty(\frac{\delta\tilde{J}}{\tilde{J}})^2 - \frac{2}{5}\delta\chi^2.\]
Below in Sec.~\ref{sec:corrected_gate_simulations}, we test the efficacy of this approach using full numerical simulations. Before we examine these results, however, we first introduce alternative approaches to suppressing noise errors.

While driving the target qubit concurrently with the cross-resonance drive is an effective strategy for suppressing the $\delta\omega_2$ error, it requires simultaneous microwave drive of both the target and control qubits, which may be impractical for some devices. As an alternative, we can use the isomorphism that exists between the $\mathfrak{su}(2)$ subalgebra of our Hamiltonian and the ordinary $\mathfrak{su}(2)$ algebra for single-qubit operations to adapt to our purposes the fastest pulse sequence that can eliminate a single-qubit drift error \cite{Zeng2018}. While Ref.~\cite{Zeng2018} assumed the ability to directly change the sign of the desired generator term, we can achieve the same effect by applying $\pi$ rotations about the $\sigma_1^z$ axis on the control qubit. Defining $\psi(\phi) \equiv \arccos(\cos(\phi / 2) / 2)$, we can correct the $\delta\omega_2$ error to lowest order for arbitrary $\phi$ with the gate sequence

\begin{align*}
	U_\phi^{\text{1Qpartial}} &= U_{\psi(\phi)-\phi/2}^{(1)} \sigma_2^z U_{2\psi(\phi) + \pi}^{(1)} \sigma_2^z U_{\psi(\phi) - \phi/2}^{(1)} \\
	&\approx e^{-i\frac{\zeta(\phi)}{2}\sigma_1^z} e^{-i \frac{\phi + \pi}{2}\sigma_1^z \sigma_2^x} \\
	&\hspace{2.5mm} + \mathcal{O}\qty(\qty(\frac{\delta\omega_2}{\tilde{J}})^2) + \mathcal{O}\qty(\frac{\delta\tilde{J}}{\tilde{J}}) + \mathcal{O}\qty(\delta\chi)
\end{align*}
where $\zeta(\phi) = \qty(4\psi(\phi) + \pi - \phi)(\eta + \delta\eta)/\tilde{J}$. If we stop here and set $\phi = \pi / 2$, we get a CNOT equivalent which is first-order insensitive to $\delta\omega_2$ errors and which never requires simultaneous drive of both qubits. In fact, using virtual gates, it should be possible to realize this gate sequence without applying any drive at all to the target qubit. This sequence, which we call ``1Qpartial," is shown in Fig.~\ref{fig:gatecircuits}c. The average gate fidelity relative to the noiseless case, to lowest order, is
\begin{align*}
	\bar{F}_{\text{1Qpartial}} &\approx 1 - 8.21 \qty(\frac{\delta\eta}{\tilde{J}})^2 - \frac{9\pi^2}{20}\qty(\frac{\delta\tilde{J}}{\tilde{J}})^2 \\
	&\hspace{2.5mm} - \frac{2}{5}\delta\chi^2 - 5.99 \qty(\frac{\delta\omega_2}{\tilde{J}})^2\frac{\delta\tilde{J}}{\tilde{J}} - 2.02 \qty(\frac{\delta\omega_2}{\tilde{J}})^4.
\end{align*}
Neglecting single-qubit gate times, which are small relative to the two-qubit gates, this gate sequence increases the total gate time by a factor of $\frac{8}{\pi}\psi(\pi/2) + 1\approx 4.08$ compared to the uncorrected CNOT.

Just as in 2Qecho, the remaining $\delta\eta$ error can be completely eliminated using $\pi$ rotations about the $\sigma_i^y$ axis on each qubit:
\begin{align*}
	U_\phi^{\text{1Qfull}} &= U_{\phi/2}^{\text{1Qpartial}} \sigma_1^y \sigma_2^y U_{\phi/2}^{\text{1Qpartial}} \sigma_1^y \sigma_2^y \\
	&\approx e^{-i\frac{\phi}{2}\sigma_1^z\sigma_2^x} + \mathcal{O}\qty(\qty(\frac{\delta\omega_2}{\tilde{J}})^2) + \mathcal{O}\qty(\frac{\delta\tilde{J}}{\tilde{J}}) + \mathcal{O}\qty(\delta\chi).
\end{align*}
The full nested gate sequence, which we call ``1Qfull," is shown in Fig.~\ref{fig:gatecircuits}d. Although 1Qfull does require driving both qubits to realize single-qubit gates, it still does not require driving both qubits simultaneously at any point. Note that, because they commute with $\sigma_1^y$, all single-qubit gates can be applied in the doubly-rotating frame using EDSR regardless of the $\chi$ rotation. In principle, single-qubit EDSR gates will also suffer some gate infidelity as a result of charge noise, reducing the fidelity of the corrected gate sequence. However, single-qubit $\pi$-pulse EDSR gate fidelities exceeding $99.9\%$ have been reported for single-electron DQD qubits \cite{Yoneda2018}, so we choose here to neglect this additional error source.

Once again setting $\phi = \pi/2$, we obtain a robust CNOT equivalent with average gate fidelity
\begin{align*}
	\bar{F}_{\text{1Qfull}} &\approx 1 - \frac{5 \pi^2}{4}\qty(\frac{\delta\tilde{J}}{\tilde{J}})^2 - \frac{2}{5}\delta\chi^2 \\
	&\hspace{2.5mm} - 20.41 \qty(\frac{\delta\omega_2}{\tilde{J}})^2\frac{\delta\tilde{J}}{\tilde{J}} - 8.44 \qty(\frac{\delta\omega_2}{\tilde{J}})^4.
\end{align*}
This additional step of correcting $\delta\eta$ errors yields a gate which is insensitive to charge noise to lowest order. However, the total time required for 1Qfull is increased by a factor $\frac{16}{\pi}\psi(\pi/4) + 3 \approx 8.55$ compared to the uncorrected CNOT, neglecting single-qubit gate times, and the effect of the $\delta\omega_2$ error has been amplified relative to 1Qpartial. For this reason, especially in the presence of accumulating error due to decoherence, it might be preferable to take advantage of the $\delta\eta = 0$ sweet spot and only correct the $\delta\omega_2$ error. This tradeoff is examined more closely in Sec.~\ref{sec:corrected_gate_simulations}, where we also provide a side-by-side comparison of the 2Qecho, 1Qpartial, and 1Qfull sequences.

\subsection{Corrected iSWAP gate}

Much of the same analysis can be applied to the iSWAP gate discussed in Refs.~\cite{Warren2019, Benito2019}. Starting with the noisy, undriven 2-qubit effective Hamiltonian with $\omega_1 = \omega_2 = \omega$,
\[H = \frac{1}{2}\qty(\omega + \delta\omega_1) \sigma_1^z + \frac{1}{2}\qty(\omega + \delta\omega_2) \sigma_2^z - \qty(J + \delta J)\sigma_1^x\sigma_2^x,\]
we move to the rotating frame for both qubits and make the RWA. In the doubly-rotating frame, we have
\begin{align*}
	H_{DF} &= \frac{1}{2}\delta\omega_1 \sigma_1^z + \frac{1}{2}\delta\omega_2 \sigma_2^z - \frac{1}{2}\qty(J+\delta J)\qty(\sigma_1^x \sigma_2^x + \sigma_1^y \sigma_2^y) \\
	&= \frac{1}{2}\delta\omega_+\frac{\sigma_1^z + \sigma_2^z}{2} + \frac{1}{2}\delta\omega_-\frac{\sigma_1^z - \sigma_2^z}{2} \\
	&\hspace{2.5mm} - \qty(J+\delta J)\frac{\sigma_1^x \sigma_2^x + \sigma_1^y \sigma_2^y}{2},
\end{align*}
where $\delta\omega_\pm = \delta\omega_1 \pm \delta\omega_2$. The gate generated by this Hamiltonian, $U_\phi = \exp(-i\frac{\phi}{2J}H_{DF})$, thus implements a noisy iSWAP local equivalent for $\phi = \pi$. The average gate fidelity relative to a noiseless iSWAP gate is
\[\bar{F} \approx 1 - \frac{\pi^2}{10}\qty(\frac{\delta J}{J})^2 - \frac{1}{10} \qty(\frac{\delta\omega_-}{J})^2 - \frac{\pi^2}{40}\qty(\frac{\delta\omega_+}{J})^2.\]
The sensitivities of the iSWAP Hamiltonian parameters to charge noise are shown in Fig.~\ref{fig:sensitivities}b. Similar to the cross-resonance gate, we find that the $\omega_i$ are much more sensitive than $J$, so we neglect the $\delta J$ error (though such errors could in principle be corrected with a more complicated gate sequence \cite{Gungordu2018}).

Thus again we have a Hamiltonian in an $\mathfrak{su}(2)\oplus\mathfrak{u}(1)$ subalgebra of $\mathfrak{su}(4)$, now with $\mathfrak{su}(2)$ generators $\{\qty(\sigma_1^x \sigma_2^x + \sigma_1^y \sigma_2^y)/2, \qty(\sigma_1^x \sigma_2^y - \sigma_1^y \sigma_2^x)/2, \qty(\sigma_1^z - \sigma_2^z)/2\}$, all of which commute with the $\mathfrak{u}(1)$ generator $\qty(\sigma_1^z + \sigma_2^z)/2$. And again, we have commuting ($\delta\omega_+$) and non-commuting ($\delta\omega_-$) error terms which we would like to eliminate. In fact, we can use the exact same nested gate sequence as for the cross-resonance gate to suppress these errors as well. Simply substituting this new $U_\phi$ for $U_\phi^{(1)}$ in the $U_\phi^{\text{1Qfull}}$ gate sequence in Fig.~\ref{fig:gatecircuits}d and setting $\phi = \pi$ yields a robust iSWAP gate which has substantially reduced sensitivity to charge noise. The fidelity of this corrected iSWAP, to lowest order, is
\[\bar{F}_{\text{1Qfull}} \approx 1 - \frac{9\pi^2}{10}\qty(\frac{\delta J}{J})^2 - 3.00 \frac{\delta J}{J}\qty(\frac{\delta\omega_-}{J})^2 - 0.25 \qty(\frac{\delta\omega_-}{J})^4.\]
Unlike the cross-resonance gate, there is no sweet spot at which the commuting error vanishes, nor can we simply suppress the non-commuting error by driving the qubits, so we must use 1Qfull to obtain a gate which corrects the largest charge noise errors. However, because the iSWAP is a $\pi$ rotation, the gate time penalty is not as severe, with 1Qfull only increasing the total gate time by a factor of $\frac{8}{\pi}\psi(\pi/2) + 1\approx 4.08$ compared to the uncorrected noisy iSWAP, neglecting single-qubit gate times.

\begin{figure}
	\input{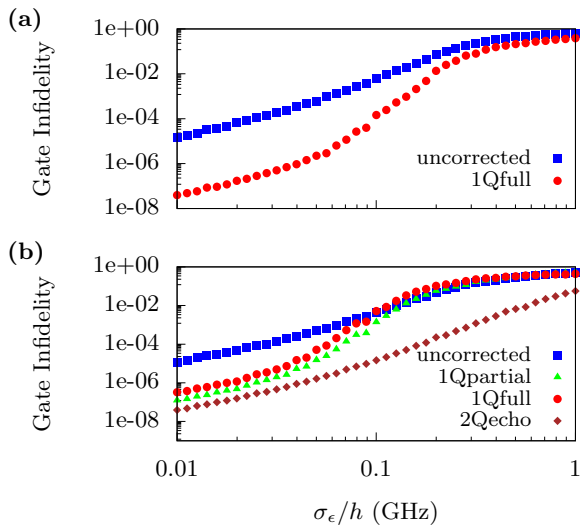}
	\caption{Average gate infidelities for various amplitudes of quasistatic charge noise, starting from the noisy doubly-rotating frame Hamiltonian. Here, we set $2 \sigma_t = \sigma_\epsilon / 100$ and have again chosen $\omega_r = \SI{6}{\giga\hertz}$, $\epsilon_1 = \epsilon_2 = 0$, $2 t_{c 1} = 2 t_{c 2} = \SI{7}{\giga\hertz}$, $g_1^{AC} = g_2^{AC} = \SI{40}{\mega\hertz}$, and $g_1^x = g_2^x = \SI{200}{\mega\hertz}$. \textbf{(a)} For the iSWAP, we choose $\omega_1 = \omega_2 = \SI{5.95}{\giga\hertz}$, and plot infidelity of the uncorrected gate $U_\pi$ as well as the corrected 1Qfull. \textbf{(b)} For the cross-resonance CNOT, we choose $\omega_1 = \SI{5.96}{\giga\hertz}$, $\omega_2 = \SI{5.94}{\giga\hertz}$, and $\Omega_1^x = \SI{28.5}{\mega\hertz}$, which tunes us to the $\delta\eta = 0$ sweet spot. Here, in addition to the uncorrected gate $U_{\pi/2}^{(1)}$ and 1Qfull, we also plot the infidelity of 1Qpartial, as well as the infidelity of 2Qecho with $\Omega_2^x = \SI{15}{\mega\hertz}$.}
	\label{fig:noiseresponse}
\end{figure}

\subsection{Corrected gate simulations}\label{sec:corrected_gate_simulations}

To investigate the effectiveness of our corrected gate sequences, we numerically compute the average gate fidelity of our corrected iSWAP and cross-resonance sequences at various quasistatic noise amplitudes. For the iSWAP (Fig.~\ref{fig:noiseresponse}a), we find that our corrected gate sequence always outperforms the uncorrected gate, at least in the absence of decohering interactions, with the improvement becoming more pronounced at smaller values of charge noise. We can roughly estimate the impact of decoherence by noting that, at short times, gate fidelity goes as $\exp(-t^2/T_2^2)$. As 1Qfull increases gate time by a factor of nearly 4 for the iSWAP, the relative penalty incurred by 1Qfull due to decoherence is then roughly $\exp(-15 t_{\text{iSWAP}}^2/T_2^2)$, where $t_{\text{iSWAP}}$ is the duration of the uncorrected iSWAP. This suggests that our gate sequence still retains its advantage if it is feasible to realize $t_{\text{iSWAP}}/T_2 \lesssim 10^{-2}$.

For the cross-resonance CNOT (Fig.~\ref{fig:noiseresponse}b), 2Qecho outperforms all other gate sequences at all levels of charge noise. As 2Qecho does not require increasing gate time beyond the addition of relatively short single-qubit gates, it also incurs no additional penalty due to decoherence. Meanwhile, 1Qfull and 1Qpartial do offer substantial fidelity improvements at sufficiently low charge noise, but no meaningful improvement is offered by either at these system parameters for $\sigma_\epsilon \gtrsim \SI{100}{\mega\hertz}$ or $\sigma_t \gtrsim \SI{1}{\mega\hertz}$. Previous experimental work in Si DQDs found detuning noise on the order of $\SI{200}{\mega\hertz}$ \cite{MiPRB2018}, suggesting that 1Qfull and 1Qpartial may offer a substantial advantage with moderate improvements over current charge noise levels, provided that gate times can be made sufficiently short relative to $T_2$. Notably, because we chose our cross-resonance pulse amplitude to tune the system to the $\delta\eta = 0$ sweet spot, 1Qpartial actually outperforms 1Qfull here. Considering also the additional penalty incurred by 1Qfull due to increased gate time, this demonstrates the considerable advantage of forgoing 1Qfull for 1Qpartial executed at the sweet spot.

\section{Conclusion}\label{sec:conclusion}

Our cross-resonance gate, with no requirement that qubits be brought into resonance, extends long-range 2-qubit entangling operations to a broader class of quantum dot architectures with a larger range of useful system parameters. With our focus on experimentally realistic parameters, we hope this work will guide efforts to develop solid-state quantum computing technologies. Additionally, our dynamic error correction sequences have the potential to greatly improve robustness to quasistatic charge noise of both our proposed cross-resonance gate as well as previously investigated cavity-mediated entangling gates, improving prospects of fault-tolerant entangling operations in solid state quantum processors.

\section*{Acknowledgments} 

It is a pleasure to acknowledge John Nichol for helpful discussions.
This work is supported by the Army Research Office (W911NF-17-0287 and W911NF-15-1-0149).

\bibliography{refs}{}

\end{document}


\title{Supplemental materials for ``Robust photon-mediated entangling gates between single-electron quantum dots"}

\author{Ada Warren}
\affiliation{Department of Physics, Virginia Tech, Blacksburg, VA 24061, USA}
\author{Utkan G\"ung\"ord\"u}
\altaffiliation{Current address: Laboratory for Physical Sciences, College Park, Maryland 20740, USA}
\affiliation{Department of Physics, University of Maryland Baltimore County, Baltimore, MD 21250, USA}
\author{J. P. Kestner}
\affiliation{Department of Physics, University of Maryland Baltimore County, Baltimore, MD 21250, USA}
\author{Edwin Barnes}
\author{Sophia E. Economou}
\affiliation{Department of Physics, Virginia Tech, Blacksburg, VA 24061, USA}

\maketitle
\onecolumngrid

In these supplemental materials, we review the derivation of the effective Hamiltonian presented in the main text and give detailed expressions for how effective Hamiltonian parameters depend on the original system parameters. Because we assume the magnetic field gradient is purely transverse, we can follow \cite{Benito2017, Benito2019} in diagonalizing the bare DQD Hamiltonian -- $\tilde{H}_0$ in the main text. We start by diagonalizing the orbital terms. Define
\begin{align*}
	\theta_i &= \arctan(\frac{2t_{c i}}{\epsilon_i}) \\
	\omega_i^a &= \sqrt{\epsilon_i^2 + 4\tau_{c i}^2} \\
	U_1 &= \prod_i\exp(-i \theta_i \tilde{\tau}_i^y / 2)
\end{align*}
and apply the unitary $U_1$ to $\tilde{H}_0$ to obtain
\[U_1^\dag \tilde{H}_0 U_1 = \sum_i \qty(\frac{1}{2}\omega_i^a\tilde{\tau}_i^z + \frac{1}{2}\omega_i^z\tilde{\sigma}_i^z + g_i^x \tilde{\sigma}_i^x\qty(\cos(\theta_i)\tilde{\tau}_i^z - \sin(\theta_i)\tilde{\tau}_i^x)).\]
Next, we can eliminate the $\tilde{\sigma}_i^x\tilde{\tau}_i^z$ terms by defining
\begin{align*}
	\alpha_i &= \arctan(\frac{2 g_i^x \cos(\theta_i)}{\omega_i^z}) \\
	\omega_i^{z'} &= \sqrt{{\omega_i^z}^2 + \qty(2 g_i^x \cos(\theta_i))^2} \\
	U_2 &= \prod_i\exp(-i \alpha_i \tilde{\tau}_i^z\tilde{\sigma}_i^y/2)
\end{align*}
and applying the unitary $U_2$ 
\[U_2^\dag U_1^\dag \tilde{H}_0 U_1 U_2 = \sum_i \qty(\frac{1}{2}\omega_i^a\tilde{\tau}_i^z + \frac{1}{2}\omega_i^{z'}\tilde{\sigma}_i^z - g_i^x \sin(\theta_i)\tilde{\sigma}_i^x\tilde{\tau}_i^x).\]
Finally, we eliminate the $\tilde{\sigma}_i^x\tilde{\tau}_i^x$ terms by defining
\begin{align*}
	\beta_i^\pm &= \arctan(\frac{-2 g_i^x \sin(\theta_i)}{\omega_i^a \pm \omega_i^{z'}}) \\
	\beta_i &= \frac{1}{2}\qty(\beta_i^+ + \beta_i^-) \\
	\omega_i^{\tau} \pm \omega_i^{\sigma} &= \sqrt{\qty(\omega_i^a \pm \omega_i^{z'})^2 + \qty(2 g_i^x \sin(\theta_i))^2} \\
	U_3 &= \prod_i \exp(-\frac{i}{2}\qty(\frac{\beta_i^+ + \beta_i^-}{2}\tilde{\tau}_i^y\tilde{\sigma}_i^x + \frac{\beta_i^+ - \beta_i^-}{2}\tilde{\tau}_i^x\tilde{\sigma}_i^y)),
\end{align*}
and now the transformation $U_1 U_2 U_3$ takes us to the eigenbasis of $\tilde{H}_0$.

Now, we define the transformed operators
\begin{align*}
	\tau_i^k &= U_1 U_2 U_3 \tilde{\tau}_i^k U_3^\dag U_2^\dag U_1^\dag \\
	\sigma_i^k &= U_1 U_2 U_3 \tilde{\sigma}_i^k U_3^\dag U_2^\dag U_1^\dag,
\end{align*}
in terms of which the system Hamiltonian becomes
\begin{align*}
	\tilde{H}_0 &= \omega_r a^\dag a + \sum_i \qty(\frac{1}{2}\omega_i^\tau \tau_i^z + \frac{1}{2}\omega_i^\sigma \sigma_i^z) \\
	\tilde{H}_I &= \sum_i g_i^{AC} \qty(a^\dag + a) d_i \\
	\tilde{H}_{dr} &= \sum_i \tilde{\Omega}_i(t) \cos(\omega_i^d t) d_i.
\end{align*}
Here, $d_i$ is the transformed electron dipole operator
\begin{align*}
d_i = \tilde{\tau}_i^z =& \frac{1}{2} \qty(\cos(\theta_i)\qty(\cos(\beta_i^+) + \cos(\beta_i^-)) - \sin(\theta_i) \sin(\alpha_i) \qty(\sin(\beta_i^+) - \sin(\beta_i^-))) \tau_i^z \\
	& + \frac{1}{2} \qty(\cos(\theta_i)\qty(\cos(\beta_i^+) - \cos(\beta_i^-)) - \sin(\theta_i) \sin(\alpha_i) \qty(\sin(\beta_i^+) + \sin(\beta_i^-))) \sigma_i^z \\
	& - \frac{1}{2} \qty(\cos(\theta_i)\qty(\sin(\beta_i^+) + \sin(\beta_i^-)) + \sin(\theta_i) \sin(\alpha_i) \qty(\cos(\beta_i^+) - \cos(\beta_i^-))) \sigma_i^x \tau_i^x \\
	& + \frac{1}{2} \qty(\cos(\theta_i)\qty(\sin(\beta_i^+) - \sin(\beta_i^-)) + \sin(\theta_i) \sin(\alpha_i) \qty(\cos(\beta_i^+) + \cos(\beta_i^-))) \sigma_i^y \tau_i^y \\
	& - \sin(\theta_i) \cos(\alpha_i) \cos(\beta_i) \tau_i^x - \sin(\theta_i) \cos(\alpha_i) \sin(\beta_i) \sigma_i^x \tau_i^z.
\end{align*}
For the special case $\epsilon_i = 0$, this takes on the much simpler form $d_i = -\cos(\beta_i) \tau_i^x - \sin(\beta_i) \sigma_i^x \tau_i^z$.

Now, we assume $\abs{\omega_i^\tau - \omega_r}, \abs{\omega_r - \omega_i^\sigma} \gg g_i^{AC}$ so that we can use a second order Schrieffer-Wolff transformation $S$ to eliminate to leading order the interactions between the DQDs and the resonator \cite{Schrieffer1966, Bravyi2011}. Furthermore, we assume that the microwave drives are sufficiently weak that we can treat $\tilde{H}_{dr}$ as first-order in the expansion, along with $H_I$. We expand the transformed Hamiltonian
\begin{align*}
e^S \tilde{H} e^{-S} &\approx \tilde{H}_0 + \tilde{H}_I + \tilde{H}_{dr} + \comm{S}{\tilde{H}_0}  + \comm{S}{\tilde{H}_I}  + \comm{S}{\tilde{H}_{dr}} + \frac{1}{2}\comm{S}{\comm{S}{\tilde{H}_0}} \\
	&= \tilde{H}_0 + \tilde{H}_{dr} + \frac{1}{2}\comm{S}{\tilde{H}_I} + \comm{S}{\tilde{H}_{dr}},
\end{align*}
where in the final line we've eliminated the leading-order DQD-resonator coupling using $\comm{S}{\tilde{H}_0} + \tilde{H}_I = 0$. This determines $S$
\begin{align*}
	S = \sum_i g_i^{AC} [&\frac{1}{2\omega_r}\qty(\cos(\theta_i)\qty(\cos(\beta_i^+) + \cos(\beta_i^-)) - \sin(\theta_i) \sin(\alpha_i) \qty(\sin(\beta_i^+) - \sin(\beta_i^-))) (a^\dag - a)\tau_i^z \\
	&+ \frac{1}{2\omega_r}\qty(\cos(\theta_i)\qty(\cos(\beta_i^+) - \cos(\beta_i^-)) - \sin(\theta_i) \sin(\alpha_i) \qty(\sin(\beta_i^+) + \sin(\beta_i^-))) (a^\dag - a)\sigma_i^z \\
	&- \qty(\sin(\theta_i)\sin(\alpha_i)\cos(\beta_i^+) + \cos(\theta_i) \sin(\beta_i^+))\qty(\frac{a^\dag \tau_i^+ \sigma_i^+ - a \tau_i^- \sigma_i^-}{\omega_r + \omega_i^\tau + \omega_i^\sigma} + \frac{a^\dag \tau_i^- \sigma_i^- - a \tau_i^+ \sigma_i^+}{\omega_r - \omega_i^\tau - \omega_i^\sigma}) \\
	&+ \qty(\sin(\theta_i)\sin(\alpha_i)\cos(\beta_i^-) - \cos(\theta_i) \sin(\beta_i^-))\qty(\frac{a^\dag \tau_i^+ \sigma_i^- - a \tau_i^- \sigma_i^+}{\omega_r + \omega_i^\tau - \omega_i^\sigma} + \frac{a^\dag \tau_i^- \sigma_i^+ - a \tau_i^+ \sigma_i^-}{\omega_r - \omega_i^\tau + \omega_i^\sigma}) \\
	&- \sin(\theta_i)\cos(\alpha_i)\cos(\beta_i)\qty(\frac{a^\dag \tau_i^+ - a \tau_i^-}{\omega_r + \omega_i^\tau} + \frac{a \tau_i^+ - a^\dag \tau_i^-}{\omega_i^\tau - \omega_r}) \\
	&- \sin(\theta_i)\cos(\alpha_i)\sin(\beta_i)\tau_i^z\qty(\frac{a^\dag \sigma_i^+ - a \sigma_i^-}{\omega_r + \omega_i^\sigma} + \frac{a \sigma_i^+ - a^\dag \sigma_i^-}{\omega_i^\sigma - \omega_r})].
\end{align*}

With all resonator couplings removed to leading order, we now assume $\abs{\omega_r - \omega_i^d} \gg \tilde{\Omega}_i(t)$ so that the microwave drives will not excite the resonator mode, and we take the empty cavity limit by projecting onto the $\ev{a^\dag a} = 0$ subspace. Additionally, we assume that $\abs{\omega_i^\tau - \omega_i^d} \gg \tilde{\Omega}_i(t)$ so that we can neglect transitions to higher-energy DQD states. This allows us to further project onto the $\ev{\tau_i^z} = -1$ subspace. Now, if we diagonalize static single-qubit terms and neglect terms higher than fourth order in $g_i^x$ or $g_i^{AC}$, we arrive at an effective Hamiltonian describing the low-energy dynamics of the DQD systems
\[H = \sum_i \qty(\frac{1}{2} \omega_i + \cos(\omega_i^d t)\qty(\Omega_i^z(t) \sigma_i^z + \Omega_i^x(t) \sigma_i^x)) - \sum_{i}\sum_{j < i} J_{ij}\sigma_i^x \sigma_j^x,\]
where we've defined
\begin{align*}
	&\omega_i = \omega_i^\sigma - 2 \omega_i^\sigma \frac{\qty(g_i^{AC})^2}{\omega_r^2 - \qty(\omega_i^\sigma)^2} \qty(\sin(\theta_i) \cos(\alpha_i) \sin(\beta_i))^2 \\
	&\hspace{5mm} + \frac{\qty(g_i^{AC})^2}{\omega_r + \omega_i^\tau + \omega_i^\sigma} \qty(\cos(\theta_i) \sin(\beta_i^+) + \sin(\theta_i) \sin(\alpha_i) \cos(\beta_i^+))^2 \\
	&\hspace{5mm} - \frac{\qty(g_i^{AC})^2}{\omega_r + \omega_i^\tau - \omega_i^\sigma} \qty(\cos(\theta_i) \sin(\beta_i^-) - \sin(\theta_i) \sin(\alpha_i) \cos(\beta_i^-))^2  \\
	&\hspace{5mm} + \frac{g_i^{AC}}{\omega_r} \qty(\cos(\theta_i)\qty(\cos(\beta_i^+) - \cos(\beta_i^-)) - \sin(\theta_i) \sin(\alpha_i)\qty(\sin(\beta_i^+) + \sin(\beta_i^-))) \\
	&\hspace{1cm} \times \sum_j g_j^{AC} \qty(\cos(\theta_j)\qty(\cos(\beta_j^+) + \cos(\beta_j^-)) - \sin(\theta_j) \sin(\alpha_j)\qty(\sin(\beta_j^+) - \sin(\beta_j^-))) \\
	&\Omega_i^x(t) = \sin(\theta_i)\cos(\alpha_i)\sin(\beta_i) \tilde{\Omega}_i(t)\\ 
	&\Omega_i^z(t) = \frac{1}{2} \qty(\cos(\theta_i)\qty(\cos(\beta_i^+) - \cos(\beta_i^-)) - \sin(\theta_i) \sin(\alpha_i)\qty(\sin(\beta_i^+) + \sin(\beta_i^-))) \tilde{\Omega}_i(t)\\
	&J_{ij} = \omega_r g_i^{AC} g_j^{AC} \sin(\theta_i)\cos(\alpha_i)\sin(\beta_i) \sin(\theta_j)\cos(\alpha_j)\sin(\beta_j) \qty(\frac{1}{\omega_r^2 - \qty(\omega_i^\sigma)^2} + \frac{1}{\omega_r^2 - \qty(\omega_j^\sigma)^2}).
\end{align*}
In the $\epsilon_i = 0$ case, this reduces to
\begin{align*}
	&\omega_i = \omega_i^\sigma - 2 \omega_i^\sigma \sin[2](\beta_i) \frac{\qty(g_i^{AC})^2}{\omega_r^2 - \qty(\omega_i^\sigma)^2} \\
	&\Omega_i^x(t) = \sin(\beta_i) \tilde{\Omega}_i(t)\\ 
	&\Omega_i^z(t) = 0\\
	&J_{ij} = \omega_r g_i^{AC} g_j^{AC} \sin(\beta_i)\sin(\beta_j) \qty(\frac{1}{\omega_r^2 - \qty(\omega_i^\sigma)^2} + \frac{1}{\omega_r^2 - \qty(\omega_j^\sigma)^2}).
\end{align*}

\bibliography{refs}{}